 \documentclass[preprint,review,12pt]{elsarticle}

\usepackage{amssymb}
\usepackage{amssymb}
\usepackage{float}
\usepackage{graphicx}
\usepackage{dcolumn}
\usepackage{amsmath}
\usepackage{subfigure}
\usepackage[hypcap]{caption}
\usepackage{hyperref}
\usepackage{epstopdf}
\usepackage{bm}
\usepackage[latin1]{inputenc}

\journal{Computers and Fluids}

\begin{document}

\begin{frontmatter}



\title{Mixing in modulated turbulence. Analytical results.}

\cortext[cor1]{email address: wouter.bos@ec-lyon.fr, tel. +33472186204}
\author{Wouter J.T. Bos\corref{cor1} 
}
\address{LMFA, CNRS, Ecole Centrale de Lyon, Universit\'e de Lyon, 
  69134 Ecully, France}

\author{Robert Rubinstein}
\address{Newport News, VA USA}

\begin{abstract}
Recent numerical results show that if a scalar is mixed by periodically forced turbulence, the average mixing rate is directly affected for  forcing frequencies small compared to the integral turbulence frequency. We elucidate this by an analytical study using simple turbulence models for spectral transfer. Adding a large amplitude modulation to the forcing of the velocity field enhances the energy transfer and simultaneously diminishes the scalar transfer. Adding a modulation to a random stirring protocol will thus negatively influence the mixing rate. We further derive the asymptotic behaviour of the response of the passive scalar quantities in the same flow for low and high forcing frequencies.    
\end{abstract}


\end{frontmatter}






\section{Introduction}

Recent results from direct numerical simulation suggest that the time-averaged mixing in turbulent flow can be affected by adding a periodic modulation to the energy injection of the turbulent velocity field \cite{Yang2016}. Indeed 
it is shown that the scalar flux towards small scales, or mixing rate, is attenuated by the modulation. On the contrary an enhanced transfer of kinetic energy towards small scales is observed. In the present work we will explain these observations  analytically using a second order analysis of simple turbulence closure models.


\subsection{Closures and turbulence}

Nowadays numerical simulations allow us to reproduce academic experiments of isotropic turbulence at Reynolds numbers of the same order as obtained in large wind-tunnel facilities \cite{Ishihara2009}. Closures have therefore partially lost their interest as tools to predict the behavior of turbulent flows at large Reynolds numbers. However, they are still very valuable for the interpretation of results on turbulent flows. If we know the minimum ingredients of a model which is capable of reproducing particular features of a turbulent flow, there is a good chance that we can obtain a qualitative understanding of the particular flow features. If we adopt this viewpoint, it is important to choose the adequate, i.e., the simplest possible model which is able to reproduce the physics of a problem. 

The simplest model for spectral transfer of  turbulent kinetic energy  is probably Kovaznay's closure \cite{Kovaznay}. It should be used if the only features one needs to retain of the nonlinear transfer is the constant flux dynamics of a long inertial range, consistent with Kolmogorov's ideas. Another simple, but more elegant closure was proposed by Leith \cite{Leith}. In addition to giving the Kolmogorov scaling, his diffusion-approximation is compatible with the thermal equilibrium properties of Galerkin truncated inviscid turbulent flow.

In reference \cite{Rubinstein2013} the simplest turbulence models were assessed for the case of scalar mixing in isotropic turbulence. Indeed, also in the scalar case, Kovaznay's and Leith's models \cite{Leith3} are examples of simple models consistent with a conserved flux and local interactions. In the present work we will use such simple closure models to explain the observations of the recent numerical study \cite{Yang2016} on mixing in periodically forced turbulence.



More precisely, the questions that we will answer in this manuscript are the following:
\begin{itemize}
 \item What is the frequency response of the scalar field at low and high frequencies  of the modulation of the velocity forcing?
 \item How is the mixing rate affected when the amplitude of the periodic part of the forcing is not small? 
\end{itemize}
 Here and in the following, a low (high) frequency refers to a frequency $\omega$ of the periodic forcing, small (large) compared to the integral frequency of the turbulent flow.

\subsection{Notation}

In the present work we consider both one-point and two-point statistics, as well as time-, phase- and ensemble-averages, and therefore a word on the notation is helpful in order to clearly distinguish the different types of statistics. 

To start with, all quantities we consider are averaged, and since we consider only statistically homogeneous flows, all quantities are independent of the space coordinate. If no tildes or overlines are used, the quantities are ensemble averages, such as $k=k(\omega, t)$, depending in principle on the frequency $\omega$ and time $t$.  We decompose the ensemble averages in a time-averaged part, indicated by overlined symbols, such as $\overline k=\overline k(\omega)$ and the remaining, periodic quantities indicated by a tilde, $\tilde k=\tilde k(\omega)$ (see for instance reference \cite{Hussain1970}). Lower-case symbols ($k,\epsilon, p$) are one-point statistics, upper-case symbols are wavenumber spectra ($E=E(\kappa,\omega,t)$, $\overline E=\overline E(\omega,\kappa), ...$, with $\kappa$ the wavenumber). Phase-shifts are indicated by $\phi=\phi(\omega)$, for the one-point statistics and $\Phi=\Phi(\kappa,\omega)$ for the wavenumber spectra.


\section{Periodically forced turbulence and mixing}

\subsection{One-point statistics}

We investigate how turbulent mixing is affected if the velocity statistics contain a periodic component. The energy balance is 
\begin{equation}\label{eq:dkdt}
 \partial_t k=p-\epsilon,
\end{equation}
where $k,p,\epsilon$ are ensemble-averages of the kinetic energy, production and dissipation, respectively. The periodic component of the velocity field is created by a modulated forcing,
\begin{equation}
 p=\overline p+\tilde p\cos(\omega t).
\end{equation}
The time-averaged and periodic components satisfy,
\begin{equation}
 0=\overline p-\overline \epsilon,
\end{equation}
and
\begin{equation}
\omega \tilde k \sin(\omega t+\phi_k)=\tilde \epsilon \cos(\omega t+\phi_{\epsilon})-\tilde p \cos(\omega t).
\end{equation}
In this expression $\phi_k$ and $\phi_\epsilon$ are phase-shifts. We have assumed that the periodic response of the velocity statistics will be dominated by a contribution at the forcing frequency $\omega$. The frequency responses of $\tilde k$ and $\tilde \epsilon$ were investigated theoretically in \cite{Bos2007-3,Heydt2003-1}, numerically in \cite{Kuczaj2006,Kuczaj2008} and experimentally in \cite{Cadot2003}.

The scalar is forced by an injection $p_\theta=\overline p_\theta$, without periodic component. But since the velocity field which advects the scalar field is modulated, a trace of the modulation can be found back in the scalar statistics. The scalar variance evolves as 
\begin{equation}
 \partial_t k_\theta= p_\theta-\epsilon_\theta.
\end{equation}
The mean and periodic parts of the scalar are given, respectively, by
\begin{equation}
0=\overline p_\theta -\overline \epsilon_\theta,
\end{equation}
and
\begin{equation}
\omega \tilde k_\theta \sin(\omega t+\phi_\theta)=\tilde \epsilon_\theta \cos(\omega t+\phi_{\epsilon_\theta}).
\end{equation}
An interesting feature of this last equation is that it does not contain any source term. On the level of single point averages it is therefore difficult to do any predictions on the behavior of $\tilde k_\theta$ and $\tilde \epsilon_\theta$.  The only result obtained from this equation is that
\begin{equation}\label{eq:omegakt}
 \omega |\tilde k_\theta|=|\tilde \epsilon_\theta|,
\end{equation}
and that the phase-shifts will differ by a constant value of $\pi/2$.

If we define the velocity and scalar timescales as
\begin{equation}
 T=\frac{\overline k}{\overline \epsilon}, ~~~~~   T_\theta=\frac{\overline k_\theta}{\overline \epsilon_\theta},
\end{equation}
then we can relate the mixing rate to the scalar timescale by 
\begin{equation}
\chi_\theta=  T_\theta^{-1}. 
\end{equation}
Equivalently we can relate the energy transfer rate to the inverse of the integral timescale,
\begin{equation}
\chi=  T^{-1}. 
\end{equation}
In the present case of statistically homogeneous flow, these definitions of the transfer and mixing rate are straightforward choices. In inhomogeneous flows the mixing process will also involve the turbulent diffusion and the definition of a mixing rate is less straightforward.  When the scalar quantity is not passive, the definitions of mixing efficiency and mixing rate are far more complicated (as for instance in stably stratified flows \cite{Winters1995}, or flows generated by Rayleigh-Taylor instabilities \cite{Lawrie2011} or Richtmyer-Meshkov instabilities \cite{Zhou2016}).

\subsection{Spectral description of the problem}

The case we consider consists in an isotropic turbulent flow, maintained statistically stationary by a forcing containing a time-periodic component. More precisely, we consider the dynamics of a turbulent flow, characterized by the kinetic energy spectrum $E(\kappa,t)$, given by the evolution equation
\begin{equation}\label{eq:dEdt}
 \partial_t E=-\partial_\kappa\Pi -2\nu \kappa^2 E+P.
\end{equation}
where 
\begin{equation}
 P=\overline P+\tilde P\cos(\omega t),
\end{equation}
and $\nu$ is the kinematic viscosity. The choice of a model for the spectral flux $\Pi$ determines the precise dynamics of the flow. An important property of this flux is that it should vanish at $k=0$ and $k=\infty$, so that the integral of  (\ref{eq:dEdt}) yields  (\ref{eq:dkdt}). 

We consider the advection of a passive scalar in the same flow, which is forced by a large scale injection, without any periodic component ($P_\theta=\overline P_\theta$). The scalar spectrum $E_\theta(\kappa,t)$ obeys,
\begin{equation}
 \partial_t E_\theta=-\partial_\kappa\Pi_\theta -2\alpha \kappa^2 E_\theta+P_\theta,
\end{equation}
where $\alpha$ is the diffusivity of the scalar. The case of unity Prandtl number, $\nu=\alpha$ is investigated. 

We consider that both the velocity and scalar forcing are confined to the largest scales. To facilitate the considerations, we assume the forcing terms to act on wavenumber $\kappa_f$ only,
\begin{equation}
 P=p~\delta(\kappa-\kappa_f),~~~~~P_\theta=p_\theta~\delta(\kappa-\kappa_f).
\end{equation}
A more complicated case could be investigated, where the periodic forcing acts at another scale than the steady part of the forcing, with possible relevance to flow-control, or geophysical flows, but this case is not considered here.

The energy spectrum and scalar spectrum will contain a time-averaged component and a periodic component,
\begin{eqnarray}
 E=\overline E+\tilde E\cos(\omega t+\Phi_E),\\
 E_\theta=\overline E_\theta+\tilde E_\theta\cos(\omega t+\Phi_{E_\theta}).
\end{eqnarray}
The time-averaged spectra obey then the equations,
\begin{eqnarray}
0= -\partial_\kappa\overline \Pi -2\nu \kappa^2 \overline E+\overline P\\
0= -\partial_\kappa\overline \Pi_\theta -2\alpha \kappa^2 \overline E_\theta+\overline P_\theta,
\end{eqnarray}
and the periodic contributions,
\begin{eqnarray}
 \left(\partial_t+2\nu \kappa^2\right) \tilde E \cos(\omega t+\Phi_{E})&=&-\partial_\kappa \left(\tilde \Pi \cos(\omega t +\Phi_\Pi)\right)+\tilde P\cos(\omega t) \label{eq:dttildeE}\\
\left(\partial_t+2\alpha \kappa^2\right) \tilde E_\theta\cos(\omega t+\Phi_{E_\theta})&=&-\partial_\kappa \left(\tilde \Pi_\theta \cos(\omega t +\Phi_{\Pi_\theta})\right).\label{eq:dttildeEt}
\end{eqnarray}
Upto this point we have not introduced any serious approximations, apart from the assumption that the response of the fields will be periodic with frequency $\omega$. In the following, section \ref{sec:Low} we will first focus on the low frequency dynamics of the velocity and scalar statistics. Then, in section \ref{sec:high}, the high frequency dynamics of the scalar will be investigated. 

\section{Low frequency dynamics \label{sec:Low}}

In the low frequency, or quasi-static limit all scales have sufficient time to adapt to the forcing. The modulation does therefore not introduce any phase-shifts in the statistics. This means that in equations (\ref{eq:dttildeE}) and (\ref{eq:dttildeEt}) the quantities $\Phi_{E}$, $\Phi_{E_\theta}$, $\Phi_{\Pi}$ and $\Phi_{\Pi_\theta}$ are close to zero. Also, the time-derivative vanishes in this limit. The equations simplify then to
\begin{eqnarray}
2\nu \kappa^2 \tilde E &=&-\partial_\kappa \tilde \Pi +\tilde P \label{eq:dttildeE-2}\\
2\alpha \kappa^2 \tilde E_\theta&=&-\partial_\kappa \tilde \Pi_\theta.\label{eq:dttildeEt-2}
\end{eqnarray}

To solve these equations, we need to specify the fluxes $\Pi$ and $\Pi_\theta$. The simplest model for those fluxes, compatible with conserved flux and Kolmogorov-Obukhov scaling, is Kovaznay's closure for the nonlinear transfer of kinetic energy \cite{Kovaznay} and its analog for scalar turbulence \cite{Rubinstein2013}. These models read 
\begin{equation}\label{eq:Kov}
 \Pi=C_k^{-3/2} \kappa^{5/2} E^{3/2},
\end{equation}
with $C_k$ the Kolmogorov constant, and
\begin{equation}\label{eq:Kovt}
 \Pi_\theta= C_\theta \kappa^{5/2}E^{1/2} E_\theta,
\end{equation}
where $C_\theta$ is related to the Kolmogorov and Corrsin-Obukhov constants by the relation
\begin{equation}
 C_\theta=(C_k C_{CO}^2)^{-1/2}.
\end{equation}
These closures are consistent with the dominant effect of straining in the transfer of energy and scalar variance towards small scales, compatible with Kolmogorov and Corrsin-Obukhov scaling. The influence of sweeping is not dominant in the energy transfer process, but its effect on the mixing efficiency constitutes an interesting direction for further research \cite{Zhou2004,Zhou2010}. 

Now these models are specified we can solve the equations to assess the influence of the modulated forcing on the dynamics of $E$ and $E_\theta$.

\subsection{Modulated kinetic energy}

Introducing $E=\overline E+\tilde E$ in (\ref{eq:Kov}) and expanding in terms of $\tilde E/\bar E$ we obtain,
\begin{equation}
\Pi\approx C_k^{-3/2} \kappa^{5/2} \overline E^{3/2}\left(1+\frac{3}{2}\frac{\tilde E \cos(\omega t)}{\overline E} +\mathcal O((\tilde E/\overline E)^2)\right), 
\end{equation}
So that 
\begin{equation}
\overline{\Pi}\approx C_k^{-3/2} \kappa^{5/2} \overline E^{3/2}\left(1+\mathcal O((\tilde E/\overline E)^2)\right), \label{eq:PiFirstOrder}
\end{equation}
and 
\begin{equation}
\tilde{\Pi}\cos(\omega t)\approx C_k^{-3/2} \kappa^{5/2} \overline E^{3/2}\left(\frac{3}{2}\frac{\tilde E \cos(\omega t)}{\overline E} +\mathcal O((\tilde E/\overline E)^2)\right).
\end{equation}
Combining these expressions, we have therefore to first order
\begin{equation}
\tilde{\Pi} =\frac{3}{2} \frac{\tilde E }{\overline E} \overline \Pi.
\end{equation}
In the inertial range, in the quasi-static limit, fluxes should balance the  energy input, 
\begin{equation}
\overline \Pi\approx \int^k \overline P dk=\overline p, ~~~~ \tilde \Pi\approx\int^k \tilde P dk=\tilde p, 
\end{equation}
which gives
\begin{equation}
\tilde{E} =\frac{2}{3} \frac{\tilde p }{\overline p} \overline E.\label{eq:tildeE23}
\end{equation}
Integration gives the modulated kinetic energy in the quasi-static limit,
\begin{equation}
\tilde{k} =\frac{2}{3} \frac{\tilde p }{\overline p} \overline k,
\end{equation}
a result which was also obtained in \cite{Bos2007-3} by elementary arguments.

\subsection{Average kinetic energy and integral timescale}

From equation (\ref{eq:PiFirstOrder}) it can be seen that, to first order, the modulated energy input does not influence the time-averaged dynamics. This is consistent with previous studies of modulated turbulence, where the amplitude of the modulated forcing was small compared to the averaged component \cite{Kuczaj2006}, \cite{Bos2007-3}. However, in the recent simulations by Yang \emph{et al.} \cite{Yang2016}, where the amplitude of the modulation was taken equal to the value of the time-averaged component, a significant modification of the averaged kinetic energy was observed for low frequencies. For such amplitudes, it is possible that the linear approximation breaks down, and higher order contributions should be taken into account. Doing so, to second order, expression (\ref{eq:PiFirstOrder}) becomes
\begin{equation}\label{eq:meansecondorder}
\overline{\Pi}= C_k^{-3/2}\kappa^{5/2} \overline{E}^{3/2}\left(1+\frac{3}{8}\frac{\overline{\tilde E^2 \cos^2(\omega t)}}{\overline {E}^2}\right).
\end{equation}
The time-average of the quadratic fluctuation is, 
\begin{equation}\label{eq:meancos2}
\overline{\tilde E^2\cos^2(\omega t)}=\frac{1}{2}\tilde E^2,
\end{equation}
and combining this with (\ref{eq:tildeE23}), we obtain in the inertial range
\begin{equation}
 \overline p=\overline \epsilon=\overline \Pi=C_k^{-3/2} \kappa^{5/2} \overline{E}^{3/2}\left(1+\frac{1}{12}\left(\frac{\tilde p}{\overline p}\right)^2 \right).
\end{equation}
This shows that the modulation leads to an enhanced energy transfer to small scales. The physical consequence of this is that if there is a time-periodic part in the large-scale forcing of a flow, the energy spectrum will be equal to 
\begin{equation}\label{eq:Ek'}
 \overline E(k)=C_k' \overline{\epsilon}^{2/3}k^{-5/3},
\end{equation}
where the Kolmogorov constant is affected by the second-order correction. Its effective value will be
\begin{equation}\label{eq:Ck'}
 C_k'=C_k\left(1+\frac{1}{12}\left(\frac{\tilde p}{\overline p}\right)^2 \right)^{-2/3},
\end{equation}
and for a fixed energy input, the kinetic energy will be lowered by the same factor, with respect to the unperturbed value $\overline k_\infty$
\begin{equation}
 \overline k=\overline k_\infty \left(1+\frac{1}{12}\left(\frac{\tilde p}{\overline p}\right)^2 \right)^{-2/3}.
\end{equation}

\subsection{Modulated scalar variance}

Introducing $E=\overline E+\tilde E$ and $E_\theta=\overline E_\theta+\tilde E_\theta$ in (\ref{eq:Kovt}) and expanding in terms of  $\tilde E/\bar E$ and $\tilde E_\theta/\bar E_\theta$ we obtain to first order
\begin{eqnarray}
 \tilde \Pi_\theta \approx 
 C_\theta\kappa^{5/2}    \overline E^{1/2} \overline E_\theta\left( \frac{\tilde E_\theta}{\overline E_\theta} + \frac{1}{2}\frac{\tilde E}{\overline E} \right). \label{eq:Pitfirstorder}
\end{eqnarray}
Since there is no modulated production term, the flux in the inertial range should vanish. This gives the relation
\begin{equation}
\frac{\tilde E_\theta}{\overline E_\theta} + \frac{1}{2}\frac{\tilde E}{\overline E}=0,
\end{equation}
and thereby
\begin{equation}
\tilde E_\theta=-\frac{1}{2}\frac{\tilde E}{\overline E}\overline E_\theta.
\end{equation}
Using (\ref{eq:tildeE23}), we find that
\begin{equation}\label{eq:tildeEt}
\tilde E_\theta=-\frac{1}{3}\frac{\tilde p}{\overline p}\overline E_\theta.
\end{equation}
The minus sign in this expression shows that the scalar modulation is always completely out of phase with respect to the kinetic energy. The modulated scalar variance is obtained by integration, yielding
\begin{equation}
|\tilde k_\theta|=\frac{1}{3}\frac{\tilde p}{\overline p}\overline k_\theta,
\end{equation}
and according to (\ref{eq:omegakt}),
\begin{equation}
|\tilde \epsilon_\theta|=\frac{1}{3}\frac{\tilde p}{\overline p}\overline \epsilon_\theta \omega.
\end{equation}

\subsection{Average scalar variance and mixing rate}

\begin{figure}
\begin{center}
\includegraphics[width=0.7\textwidth]{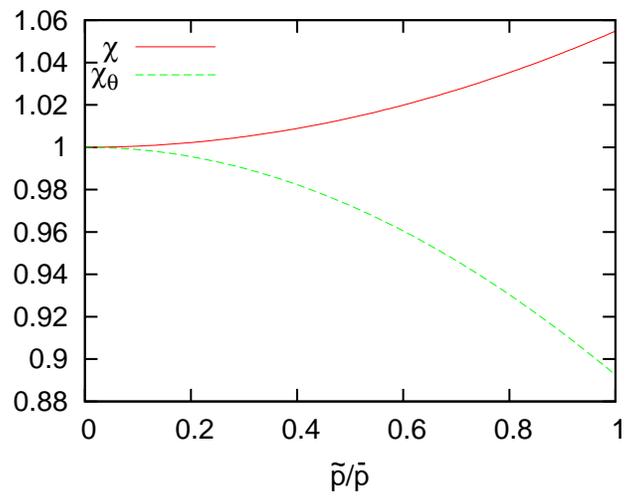}
\caption{Transfer rate $\chi$ and mixing rate $\chi_\theta$ in the low frequency limit, as a function of the relative forcing amplitude. Adding a slow, large amplitude modulation to the energy input diminishes the mixing rate and enhances the kinetic energy transfer rate.\label{Fig:1}} 
\end{center}
\end{figure}

Again, as for the average kinetic energy, the modulation of the flow does not affect the linear expansion of the average scalar flux $\overline \Pi_\theta$. Retaining second order contributions in the expansion of (\ref{eq:Kovt}), we obtain
\begin{equation}
\overline \Pi_\theta=C_\theta \kappa^{5/2}\overline E_\theta \overline E^{1/2}   \left( 1+\frac{1}{2} \overline{\frac{\tilde E_\theta \tilde E  }{\overline E_\theta \overline E}\cos^2(\omega t)} -\frac{1}{8}  \overline{\frac{\tilde E^2}{\overline E^2}\cos^2(\omega t)}          \right)
\end{equation}
using (\ref{eq:tildeEt}), (\ref{eq:meancos2}) and (\ref{eq:tildeE23})  this gives
\begin{equation}
\overline \Pi_\theta=C_\theta \kappa^{5/2}\overline E_\theta \overline E^{1/2}   \left( 1- \frac{1}{12}\left(\frac{\tilde p}{\overline p}\right)^2        \right).
\end{equation}
Again, as for the kinetic energy, we see that the modulation affects the flux, but this time the effect is the opposite: the transfer rate is diminished. For a fixed input, in the inertial range we have,
\begin{equation}
 \overline p_\theta=\overline \epsilon_\theta=\overline \Pi_\theta, 
\end{equation}
yielding, with (\ref{eq:Ek'}) and \ref{eq:Ck'},
\begin{equation}
 E_\theta=C'_{CO}\overline \epsilon_\theta \overline \epsilon^{-1/3}k^{-5/3},
\end{equation}
where 
\begin{equation}\label{eq:CCO'}
 C_{CO}'=C_{CO}\frac{\left(1+\frac{1}{12}\left(\frac{\tilde p}{\overline p}\right)^2 \right)^{1/3}}{\left( 1- \frac{1}{12}\left(\frac{\tilde p}{\overline p}\right)^2        \right)}.
\end{equation}
The variance of the passive scalar is then enhanced by the same factor,
\begin{equation}
 \overline k_\theta=\overline k_\theta(\tilde p=0)\frac{C'_{CO}}{C_{CO}}.
\end{equation}
Since the mixing rate can be defined as
\begin{equation}
 \chi_\theta= \overline \epsilon_\theta/\overline k_\theta,
\end{equation}
we have for a fixed scalar input,
\begin{equation}
\chi_\theta(\tilde p)=\chi_\theta(\tilde p=0)\frac{C_{CO}}{C'_{CO}}. 
\end{equation}
This expression is shown in Figure \ref{Fig:1}, together with the kinetic energy transfer rate. The mixing rate goes down for a large-amplitude modulation and the energy transfer rate increases, as observed in the DNS \cite{Yang2016}.

\section{High frequency behavior of the modulated variances\label{sec:high}}

In the previous section, the quasi-static limit was considered. In that limit the different phase-shifts disappeared from the dynamics. When the frequency goes up, the phase-shifts $\Phi_E$ and $\Phi_{E_\theta}$, both functions of $\omega$ and $\kappa$, complicate the analysis. In \cite{Bos2007-3}, it was shown that it is convenient to rewrite these equations in terms of spectra in- and out-of-phase, 
\begin{eqnarray}
 E=\overline E+\tilde F\cos(\omega t)+\tilde G\sin(\omega t), \label{eq:FG}\\
 E_\theta=\overline E_\theta+\tilde F_\theta\cos(\omega t)+\tilde G_\theta\sin(\omega t). \label{eq:FGt}
\end{eqnarray}
We have hereby eliminated the phases $\Phi_E$ and $\Phi_{E_\theta}$ from our description. Both descriptions are obviously equivalent and can be related by
\begin{equation}
\tilde F^2+\tilde G^2=\tilde E^2~~~~\textrm{and}~~~~\tan(\Phi_E)=-\tilde G/\tilde F, 
\end{equation}
with equivalent expressions for the scalar case.
The first question to address is how the quantities $\tilde F_\theta$ and $\tilde G_\theta$ will be influenced by the periodically fluctuating energy input. The expressions for $\tilde F$ and $\tilde G$ were determined in \cite{Bos2007-3}. It was shown that in the inertial range, for large $\omega$, $\tilde F,\tilde G$ were given by
\begin{equation}\label{eq:FGBos2007}
 \tilde G /\overline E\sim \omega^{-3}\kappa_f^2 \tilde p~~~\textrm{and}~~~  \tilde F/\overline E\sim \omega^{-2}\kappa_f^2 \tilde p\epsilon^{-1/3}\kappa^{-2/3}.
\end{equation}
In the light of our recent investigation of out-of-equilibrium turbulence \cite{Bos2016}, $\tilde G$ corresponds to the equilibrium and $\tilde F$ to the non-equilibrium contribution of the modulated kinetic energy spectrum.

\begin{figure}
\begin{center}
\includegraphics[width=0.7\textwidth]{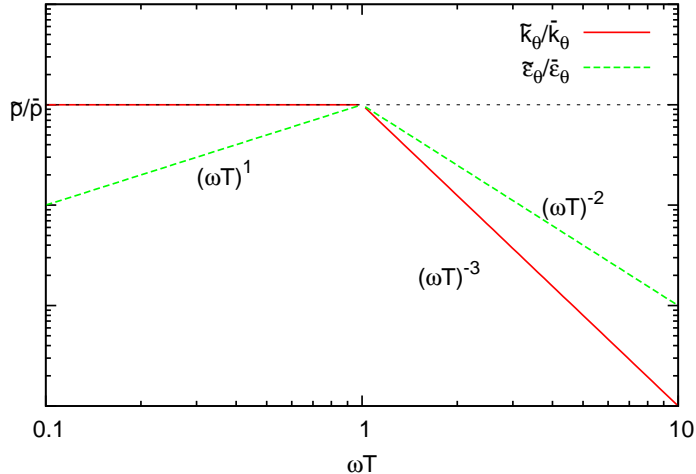}
\caption{Frequency response of the modulated part of the passive scalar variance and dissipation. \label{Fig:2}} 
\end{center}
\end{figure}

The derivation of the large-frequency behaviour of $\tilde k_\theta$ is straightforward. First we replace in $\Pi_\theta$  (expression (\ref{eq:Kovt})) the energy spectrum and the scalar spectrum by their expressions (\ref{eq:FG}) and (\ref{eq:FGt}). Assuming the fluctuating spectra small compared to the time-averaged spectra, $\tilde E/\overline{E}\ll 1, \tilde E_\theta/\overline{E}_\theta\ll 1$, the flux can be approximated (to first order) by,
\begin{eqnarray}
 \tilde \Pi_\theta \cos(\omega t +\phi_\Pi)\approx 
 \overline \Pi_\theta\left(\frac{\tilde F_\theta \cos(\omega t)+ \tilde G_\theta \sin(\omega t)}{\overline E_\theta}+\frac{1}{2}\frac{\tilde F\cos(\omega t)+\tilde G\sin(\omega t)}{\overline E} \right).
\end{eqnarray}
We substitute this expression in the equation for the fluctuating scalar spectrum (\ref{eq:dttildeEt}). Evaluating the expression at $t=0$ and $t=\pi/2$ yields two coupled equations,
\begin{eqnarray}
 -\omega \tilde F_\theta=-\partial_\kappa\left(\overline \Pi_\theta\left(\frac{\tilde G_\theta}{\overline E_\theta} +\frac{1}{2}\frac{\tilde G}{\overline E}\right)\right)-2\alpha \kappa^2 \tilde G_\theta   \label{eq:Ftheta}\\
 \omega \tilde G_\theta=-\partial_\kappa\left(\overline \Pi_\theta\left(\frac{\tilde F_\theta}{\overline E_\theta} +\frac{1}{2}\frac{\tilde F}{\overline E}\right)\right) 
 -2\alpha \kappa^2 \tilde F_\theta.\label{eq:Gtheta}
\end{eqnarray}
In the inertial range, we assume that
\begin{equation}
 \overline E(k)=C_k \overline\epsilon^{2/3}\kappa^{-5/3},~~~~~~ E_\theta=C_{CO} \overline \epsilon_\theta \overline\epsilon^{-1/3}\kappa^{-5/3}.
\end{equation}
We further use the results (\ref{eq:FGBos2007}), and, ignoring the diffusive terms, the above set of equations can be rewritten as
\begin{equation}
\frac{\tilde F_\theta}{\overline E_\theta}= \psi\partial_\kappa\left(\frac{\tilde G_\theta}{\overline E_\theta}\right)
\end{equation}
and
\begin{equation}\label{eq:[G]}
\left[1+\psi\partial_\kappa\left[\psi\partial_\kappa\right]\right] \frac{\tilde G_\theta}{\overline E_\theta}= \psi\omega^{-2}\kappa_f^2 \tilde p\overline \epsilon^{-1/3}\kappa^{-5/3}    
\end{equation}
where we introduced the expression,
\begin{equation}
 \psi\sim \omega^{-1}\overline \epsilon^{1/3}k^{5/3}.
\end{equation}
The second term in (\ref{eq:[G]}), containing the second derivative, is proportional to $\omega^{-2}$. For large frequencies we can thus ignore it compared to the first one, yielding,
\begin{eqnarray}
\frac{\tilde G_\theta}{\overline E_\theta}\sim \omega^{-3}\kappa_f^2 \tilde p
\end{eqnarray}
and 
\begin{eqnarray}
\tilde F_\theta\sim \frac{\epsilon^{1/3}k^{2/3}}{\omega}\tilde G_\theta.
\end{eqnarray}
Since $\bar p=\bar \epsilon\sim \bar k/T$ and $\kappa_f\sim \bar \epsilon \bar k^{-3/2}$, this can be expressed as
\begin{eqnarray}
\frac{\tilde G_\theta}{\overline E_\theta}\sim (\omega T)^{-3} (\tilde p/\bar p), 
\end{eqnarray}
and therefore
\begin{eqnarray}
\tilde k_\theta \sim \bar k_\theta (\omega T)^{-3} (\tilde p/\bar p).
\end{eqnarray}
We see thus that the periodic part of the scalar variance, at large $\omega$ is determined by large scale quantities. Furthermore, the scalar dissipation is given directly by
\begin{equation} 
|\tilde \epsilon_\theta|=\omega |\tilde k_\theta|,
\end{equation}
and therefore,
\begin{equation}
 \tilde \epsilon_\theta=\frac{\tilde p}{\overline p}\overline \epsilon_\theta(\omega T)^{-2}.
\end{equation}
Our estimates for the low and high frequency asymptotes are sketched in Figure \ref{Fig:2}.

\section{Discussion}

The main result obtained in the present manuscript involves the mixing rate. According to Figure 1, the mixing rate $\chi_\theta$ will go down to $89\%$ of its value if the flow-forcing contains a modulation with an amplitude ratio $\tilde p/\overline p=1$. The energy-transfer rate, on the other hand will increase towards $\chi/\chi(\tilde p=0)=1.05$. These figures are the same order of magnitude as the figures obtained in the numerical study \cite{Yang2016}. Note that at an amplitude of $\frac{\tilde p}{\overline p}=0.2$ the change in the mixing rate and energy transfer becomes less than $1\%$. It is thus clearly a large amplitude effect. For smaller forcing amplitudes the linear prediction that the average fields are unaffected is approximately satisfied.

The same conclusions hold for the Kolmogorov and Corrsin-Obukhov constant. Their variations also become important only for large amplitude ratios. A recent related study of a similar effect can be found in \cite{Chien2013}. They exploited experimentally the idea by Monin and Yaglom \cite{Monin}, section 25.1, who introduced a model to illustrate the influence of a fluctuating dissipation rate, or energy injection rate, on the Kolmogorov constant.


We can summarize the new physical insights as follows. The time-averaged nonlinear transfer is, at second order, influenced by the modulation of the energy injection. This enhances the energy flux. For a given energy injection at a given scale, this will decrease the level of the kinetic energy, or equivalently, the value of the Kolmogorov constant. The transfer of the passive scalar is also affected by the modulation, but the influence is the contrary. Qualitatively, this can be understood, at least partially, by the fact that the kinetic energy is decreased, so that the mixing of the scalar by the velocity fluctuations is less efficient. Adding a modulation to the forcing of a turbulent flow will thus decrease the integral velocity timescale and, simultaneously, increase the scalar timescale.

\section*{References}



\end{document}